\documentclass[journal]{IEEEtran}

\usepackage{latexsym}
\usepackage{graphicx}
\usepackage{amsfonts,amssymb,amsmath}
\usepackage{ctable} 
\usepackage{mathtools, cuted}
\usepackage{hyperref}
\usepackage[ruled,linesnumbered,noend]{algorithm2e}
\usepackage[T1]{fontenc}
\usepackage{cite}
\usepackage{subcaption}
\usepackage{comment}
\usepackage{diagbox}

\usepackage{amsthm}
\usepackage{overpic}
\usepackage{steinmetz}
\usepackage{array}
\usepackage{url}
\usepackage{color}

\usepackage{algorithmicx}
\usepackage{algcompatible}

\usepackage{xcolor}
\usepackage{soul}

\usepackage{linegoal}

\theoremstyle{plain}

\SetCommentSty{mycommfont}

\IEEEoverridecommandlockouts

\begin{document}
\bstctlcite{BSTcontrol}
\title{Swarm and Evolutionary Computation for Near-Field Localization}

\author{
Parisa Ramezani,  Seyed Jalaleddin Mousavirad, 
 Mattias O'Nils,  and
 Emil Bj\"ornson
 \thanks{  P. Ramezani is with University of Technology Sydney, Sydney, Australia; S.J. Mousavirad and M. O'Nils are with Mid Sweden University, Sundsvall, Sweden; E. Bj\"ornson is with KTH Royal Institute of Technology, Stockholm, Sweden.
}
 }

\maketitle
\begin{abstract}
Near-field localization has attracted significant attention in recent years due to the move toward higher frequencies and extremely large aperture arrays, which expand the near-field region and bring many sources into it. This implies that antenna arrays can be used to localize not only in angle but also in range. Although a wide range of localization methods has been developed, each comes with limitations that may hinder practical deployment. This article focuses on a class of techniques that has received relatively little attention in the prior literature despite its strong potential for accurate and efficient location estimation: swarm and evolutionary computation (SEC). These methods are well-suited to the complex optimization landscape of near-field localization and can offer important advantages over conventional approaches such as grid-based subspace methods and deep learning approaches. 
\end{abstract}

\section{Introduction}
Near-field localization has become increasingly important with the move toward 6G-and-beyond wireless systems because of larger array apertures and often higher carrier frequencies. In these settings, the near-field region is no longer negligible, and the received signals depend on both the direction and the range of the sources, providing the basis for more precise localization than in traditional far-field scenarios \cite{Lu2024Near,Ramezani2025localization,Wang2025Near}.
This is particularly important for applications such as robotics, autonomous vehicles, unmanned aerial vehicle (UAV) networks, and immersive services, all of which rely on accurate and reliable position information.

Near-field localization, however, brings its own challenges. Since the received signals depend on both angle and range of the sources, the underlying signal model becomes more complicated than in far-field settings, which makes the resulting estimation problems more nonlinear. 
There has been substantial effort to address the near-field localization problem in the literature. Existing approaches include grid-based subspace techniques such as MUltiple SIgnal Classification (MUSIC) and Estimation of Signal Parameters via Rotational Invariance Techniques (ESPRIT) \cite{He2013Near, Alva2023parametric}, maximum likelihood methods \cite{Yan2025RIS}, range-based schemes using received signal strength (RSS) measurements\cite{Huang2022Near}, channel state information (CSI)-based methods \cite{Zheng2026Near}, and deep learning methods \cite{Liu2019Deep,Chen2025robust}. 

Each near-field localization method has its own limitations. Maximum likelihood approaches can provide accurate estimates, but their computational complexity is prohibitive for practical implementations. RSS-based methods are sensitive to shadowing and environmental variations, and suffer from low spatial resolution. CSI-based methods usually require fine-grained channel information. Subspace-based methods rely on discretized grids where improving accuracy requires increasingly dense grids. Deep learning methods are sensitive to the data setting on which they are trained, and their performance degrades when the operating conditions differ from those used during training.

 In this article, we focus on swarm and evolutionary computation (SEC), a class of methods that has received limited attention in the near-field localization literature. SEC methods are population-based optimization techniques that iteratively improve a set of candidate solutions by exploring the continuous search space and gradually concentrating the search around promising regions. This makes them particularly attractive for highly nonlinear localization problems with complex objective landscapes, as multiple regions of the search space can be explored simultaneously, improving robustness against poor local solutions.

SEC methods have found applications in a wide range of fields, but their use in localization problems has been relatively limited to some conventional far-field settings, such as evolutionary optimization-based localization in wireless sensor networks \cite{Salehinejad20143D} and swarm intelligence-based localization in UAV networks \cite{Arafat2019Localization}.
Recently, we used an evolutionary computation framework to estimate the locations of multiple sources located in the near-field region of an antenna array and showed that it can achieve promising performance gains over MUSIC and its low-complexity variant \cite{mousavirad2026primer}.  Here, we extend this line of investigation and provide a more comprehensive discussion of SEC-based near-field localization.
We first introduce the basics of SEC and discuss its two main components: swarm computation and evolutionary computation, using particle swarm optimization (PSO) and differential evolution (DE) as two well-established representative examples. 
We then discuss SEC for near-field localization and compare the performance of PSO and DE in different localization scenarios.  
We further present the limitations of two prominent classes of near-field localization methods, namely grid-based subspace methods and deep learning methods, and explain how SEC can help address these limitations. Finally, we identify several promising research directions
that may serve as useful guidance for researchers pursuing future work in this area.

\section{Swarm and Evolutionary Computation}
\begin{figure}
    \centering
    \includegraphics[width=0.9\linewidth]{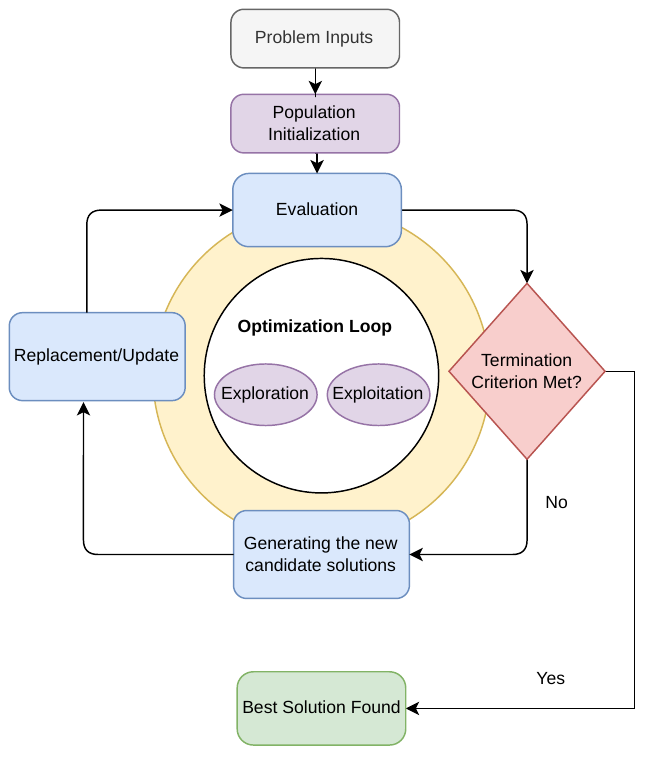}
    \caption{General scheme of SEC algorithms for solving complex multi-dimensional optimization problems.}
\label{fig:general_scheme}
\end{figure}
Many engineering problems 
have complex objective landscapes with multiple local peaks and valleys. As a result, traditional optimization methods, particularly those relying on local or gradient-based search, may struggle to converge to the correct solution.
SEC methods address such problems by maintaining a population of candidate solutions that explore the search space simultaneously, rather than refining a single point. Through iterative update rules, the population is guided toward promising regions, enabling multiple areas of the objective landscape to be explored at once. This is particularly useful for near-field localization, where nonlinear signal models can create many misleading local solutions.

The general scheme of SEC algorithms is shown in Fig.~\ref{fig:general_scheme}. These algorithms take an objective function and a feasible search region as inputs, then start with a set of randomly generated candidate solutions. Through an iterative process, the candidate solutions evolve using specific operators that balance exploration, which searches new regions of the solution space, and exploitation, which refines promising solutions locally. SEC methods typically retain the best candidate identified during the search, ensuring that the best objective value is improved or preserved over iterations. After the stopping criterion is met, this retained candidate is returned as the final solution.
Although evolutionary and swarm computation share this population-based perspective, they are inspired by different natural processes and therefore exhibit distinct search behaviors. 
In the following, we take a closer look at these two major classes of population-based methods. 

\subsection{Swarm Computation}

Swarm-based methods are motivated by the collective behavior seen in nature, such as bird flocks and fish schools. In these natural collective systems, each individual follows simple rules and shares information with others. This process helps the group make coordinated decisions with no need for any centralized control.  When applied to optimization problems, this idea is translated into a group of candidate solutions moving through the search space and sharing information about promising regions. Over time, this shared knowledge leads to high-quality solutions.

A representative example of this approach is PSO~\cite{kennedy1995particle}, where each candidate solution, referred to as a particle, updates its position by combining its own experience with information obtained from the rest of the population. In particular, each particle is influenced by the best solution it has found so far as well as the best solution identified by the entire group. This mechanism enables the swarm to quickly concentrate around promising regions of the search space. 

Fig.~\ref{fig:meta} illustrates an example search landscape for a minimization problem. As shown in the figure, the reference point corresponds to the global optimum. A typical particle moves toward a new position by combining the directions toward personal best (pbest) and global best (gbest), effectively guiding the search toward high-quality regions of the objective landscape.

\subsection{Evolutionary Computation}
Evolutionary methods are motivated by the biological principles of natural selection. Starting from a population of candidate solutions, the candidate set gradually improves through processes such as variation and selection. At each stage, new candidates are created by altering or combining existing ones, and the ones that improve the objective function are retained. This process brings gradual improvement while maintaining diversity within the population and allowing the algorithm to continue exploring new regions of the search space.

DE~\cite{storn1995differential} is a widely-used evolutionary computation method in which new candidates are generated by combining differences between existing population members. This mechanism promotes exploration because it allows the algorithm to move in directions other than the one pointing toward the current best solution. At the same time, a selection step ensures that only improved candidates are kept, gradually guiding the population toward better solutions.

\begin{figure*}
    \centering
    \includegraphics[width=0.86\linewidth]{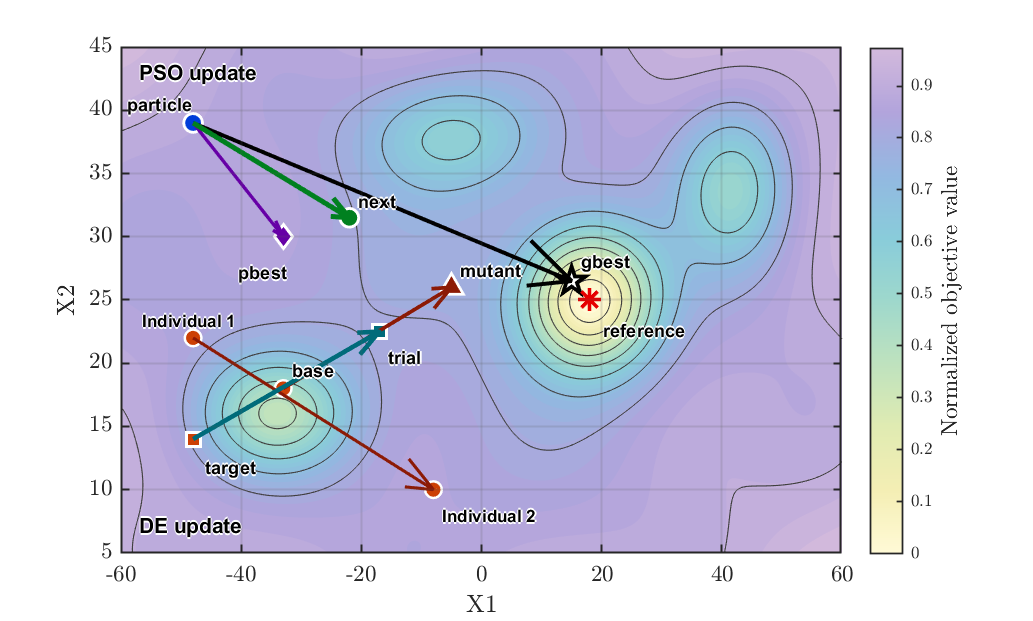}
    \caption{Conceptual illustration of PSO and DE updates on an example multimodal search landscape for a minimization problem.}
\label{fig:meta}
\end{figure*}

Fig.~\ref{fig:meta} illustrates this process on the same multimodal search landscape considered for PSO. 
A base vector and two other individuals from the population are used to generate a mutant vector, where the difference between the two individuals is added to the base vector.
The resulting mutant represents a directional exploration of the search space. This mutant is then recombined with the target solution to form a trial vector. If the trial yields a better objective value, it replaces the target in the next iteration.
Unlike PSO, where particles are attracted toward the best-known solutions, DE relies on difference-based updates between population members, which helps promote diversity during the search.

\section{SEC for Near-Field Localization}

When applying SEC methods to near-field localization, an essential first step is to cast the localization task as an optimization problem. This requires careful design choices to ensure that the resulting problem  represents the underlying physical model and leads to reliable solutions. 
Here, we focus on joint localization of multiple sources, where the number of sources is assumed to be known.

Three key components must be defined. First, each candidate solution must represent the design variables, i.e., the unknown source locations (e.g., ranges and angles, or Cartesian coordinates). Second, an objective function must be designed to evaluate how well each candidate explains the observed measurements. Third, appropriate update operators based on the chosen search strategy must be specified to guide the search process. Among these three components, the objective function plays a central role, as it shapes the optimization landscape.

\subsection{Representation and Objective Functions}

For a scenario with $K$ sources, a candidate solution typically contains the parameters of all sources and can be expressed as \begin{equation} \mathbf{x} = [\boldsymbol{\theta}_1^{T},\boldsymbol{\theta}_2^{T},\ldots,\boldsymbol{\theta}_K^{T}]^{T}, \label{eq:representation} \end{equation} where $\boldsymbol{\theta}_k$ denotes the parameter vector of the $k$-th source. In two-dimensional near-field localization, $\boldsymbol{\theta}_k=[\phi_k,r_k]^T$ contains the azimuth angle and range, whereas in three-dimensional localization it includes the elevation angle as well. Consequently, the dimensionality of the search space grows with the number of sources and the number of parameters used to describe each source.

 As SEC methods do not rely on gradients, they can work with a wide range of objective functions; however,  their success depends heavily on how well the objective function reflects the true localization goal.
The objective function should satisfy two key properties. First, it should be \emph{consistent}, meaning that its optimum corresponds to the true source locations under ideal conditions. Second, it should be \emph{discriminative}, providing sufficient contrast between correct and incorrect candidate solutions to effectively guide the search process.

Here, we introduce two different categories for the objective function. The first category includes data-fitting objectives, which directly measure how well a hypothesized set of locations represents the received measurements. The second category consists of subspace-based objectives, which rely on the eigenspace structure of the sample covariance matrix rather than fitting the received signals directly \cite{mousavirad2026primer}. 

 For the first category, a common example is a least squares-based objective functions as
 
\begin{equation}
J_{\mathrm{LS}}(\mathbf{x}) = \left\| \left(\mathbf{I}-\mathbf{P}_{\mathbf{A}(\mathbf{x})}\right) \mathbf{Y} \right\|_{F}^{2}, \label{eq:LS_obj} 
\end{equation} 
where $\mathbf{Y}$ denotes the received data matrix, $\mathbf{A}(\mathbf{x})$ is the array response matrix associated with the candidate solution, and $\mathbf{P}_{\mathbf{A}(\mathbf{x})}$ is the projection matrix onto the column space of $\mathbf{A}(\mathbf{x})$. It evaluates how well the received signals at the antenna array can be represented by the array responses associated with a candidate set of source locations. 
 
 For the second category, one option is signal-subspace fitting, which compares the subspace predicted by a hypothesized set of locations with the signal subspace estimated from the sample covariance matrix of the received data. 
 A representative formulation is given by
 
 \begin{equation}
J_{\mathrm{SF}}(\mathbf{x})
=
\left\|
\left(\mathbf{I}-\mathbf{P}_{\mathbf{A}(\mathbf{x})}\right)
\mathbf{U}_{s}
\right\|_{F}^{2},
\label{eq:SF_obj}
\end{equation}
where $\mathbf{U}_{s}$ denotes the estimated signal subspace. 
Similar formulations can also be constructed using the estimated noise subspace, leading to objective functions that resemble those employed in MUSIC-type methods.

In the evaluations presented in this article, we use a signal-subspace fitting objective function. 

\subsection{Comparative Performance of PSO and DE}

We next compare the performance of PSO and DE as representative methods of swarm and evolutionary computation, respectively, in a range of near-field localization settings. The goal is to highlight how their different search behaviors affect performance under different conditions. The relative effectiveness of PSO and DE for near-field localization is highly problem-dependent. There is no universal rule stating that one method is always superior, since their performance can change significantly with the problem setting. Nevertheless, our numerical studies have revealed several useful trends regarding the scenarios in which each method tends to be more effective, as discussed next.

We consider a $128$-element uniform linear array with half-wavelength antenna spacing as the anchor node at the carrier frequency of $30\,$GHz. The channel between the anchor node and the sources is modeled by Rician fading, with a Rician factor of $10\,$dB. The goal is to localize three sources located in the radiative near-field region of the antenna array. The ranges and angle of arrivals (AoAs) of the sources are picked from the uniform distributions $\mathcal{U}[2D,d_{\mathrm{F}}/2]$ and $\mathcal{U}[-60^\circ,60^\circ]$, respectively, where $D$ is the array aperture and $d_{\mathrm{F}}$ is the Fraunhofer distance.  
Since both PSO and DE are stochastic optimizers, we allow multiple restarts, where each restart corresponds to running the algorithm from a different random initialization, and the best solution is retained as the final solution.
For both algorithms, the population size is $80$ and the maximum number of iterations is set to $300$.
It is worth mentioning that PSO and DE are compared under the same computational budget. as the computational complexity of both algorithms is dominated by the maximum number of iterations, population size, and objective function evaluation cost, all of which are set to identical values in both schemes. 

\subsubsection{PSO and DE for Different Numbers of Sources}
\begin{figure}
    \centering
    \includegraphics[width=0.9\linewidth]{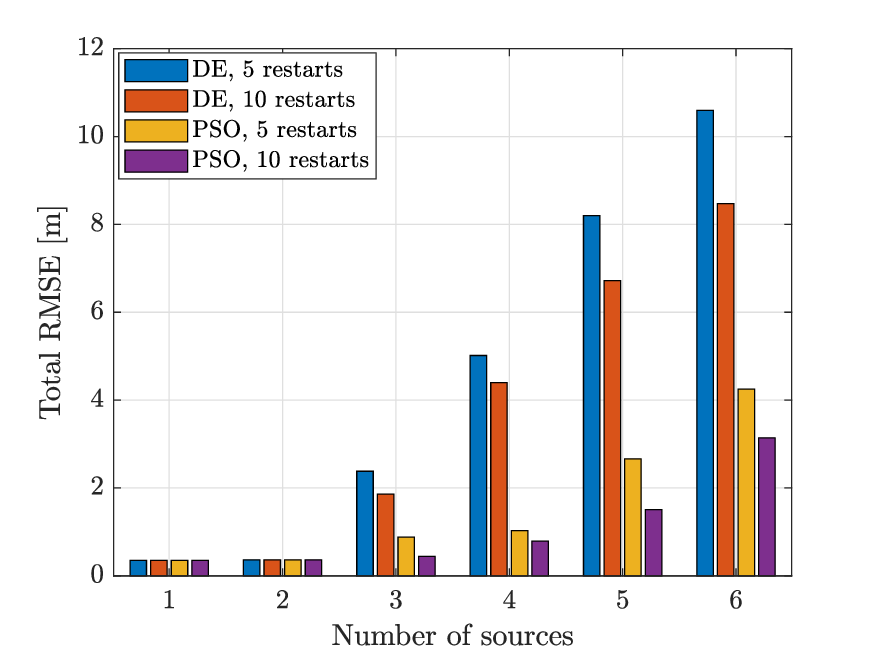}
    \caption{Performance comparison between DE and PSO for different number of sources in near-field localization.  }
\label{fig:DEvsPSO1}
\end{figure}

Fig.~\ref{fig:DEvsPSO1} shows the root mean square error (RMSE) of the location estimation for PSO and DE, computed based on the $x$-$y$ Cartesian coordinate, as a function of the number of sources. 
Two restart settings are considered: $5$ restarts and $10$ restarts. 
Increasing the number of restarts lowers the RMSE as it increases the chance that at least one run reaches a neighborhood sufficiently close to the true parameters. As a result, the RMSE improves with a larger number of restarts. 
This improvement comes at the cost of higher runtime, since each restart requires an additional full run of the optimizer. In practice, the number of restarts should be chosen to balance computational complexity against the desired accuracy and reliability.

Another important observation from Fig.~\ref{fig:DEvsPSO1} is that when the number of sources is small, both DE and PSO can converge to a good solution, so their RMSEs are similar. As the number of sources increases, DE is more likely to settle into a solution that is only partially correct. In other words, it may localize several sources well, but estimate the location of one noticeably wrong.
This behavior can be explained by the different update mechanisms of DE and PSO. In DE, a trial solution is accepted only if it immediately improves the objective value. Therefore, when most sources are already well estimated, correcting a remaining misplaced source may require a larger joint change that does not reduce the cost right away. Moreover, as the DE population becomes more similar over iterations, the mutation steps become smaller, making it harder to move a misplaced source toward the correct region. In contrast, PSO uses both personal best and global best information, which can guide particles toward a more consistent joint solution. Hence, for a larger number of well-separated sources, PSO may converge more reliably than DE.

\begin{figure}
    \centering
    \includegraphics[width=0.9\linewidth]{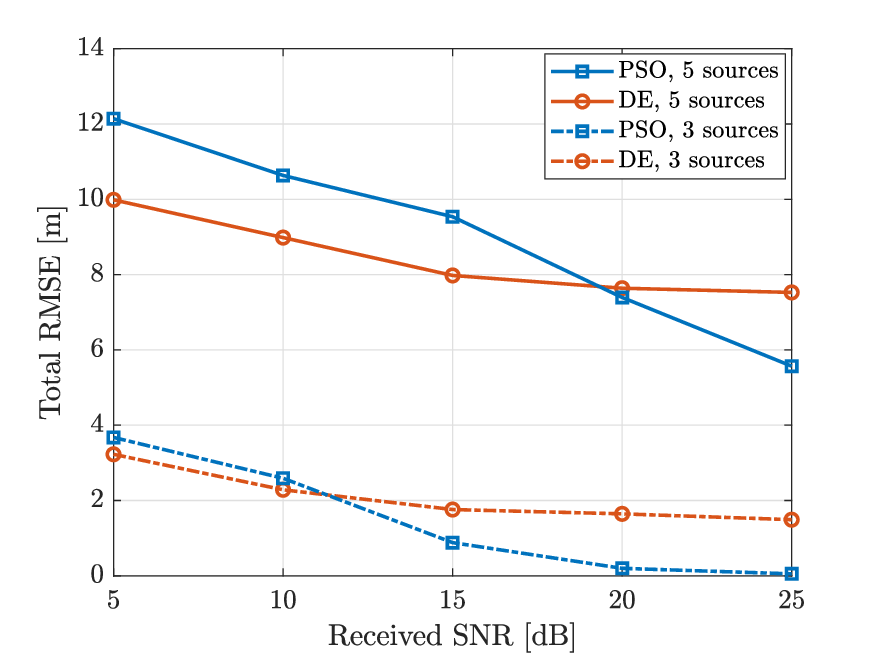}
    \caption{Performance comparison between DE and PSO for tightly-spaced source scenarios.  }
\label{fig:DEvsPSO2}
\end{figure}

\subsubsection{PSO and DE in Tightly-Spaced Source Scenarios}
Fig.~\ref{fig:DEvsPSO2} compares the performance of PSO and DE in tightly-spaced source scenarios. Two localization cases are considered, corresponding to three and five sources. In both cases, all sources are placed at the same range of $20\,\mathrm{m}$ so that the effect of tight angular spacing can be examined in isolation. For the three-source case, the AoAs are $35^\circ$, $40^\circ$, and $45^\circ$, while for the five-source case they are $35^\circ$, $40^\circ$, $45^\circ$, $50^\circ$, and $55^\circ$.

As can be seen in Fig.~\ref{fig:DEvsPSO2}, DE provides lower RMSE than PSO in the low-SNR regime. This can be attributed to the increased ambiguity of the objective function when the sources are closely spaced and the SNR is low. Under these conditions, noise and source proximity can create misleading regions in the search space and increase the risk of premature convergence.  PSO can be more susceptible to such misleading regions because particle updates are influenced by the personal best and global best positions. If some particles record misleading personal best positions, or if the global best solution lies near an incorrect region, the swarm may be pulled toward that region too early, reducing diversity and limiting further exploration.
By contrast, DE generates new candidate solutions through differences between population members and accepts them only when they improve the objective value. This mutation-selection mechanism preserves exploration more effectively and prevents the population from collapsing too quickly around an incorrect solution.
 As the SNR increases, the underlying structure of the objective function becomes more clear, allowing PSO to converge more effectively once the correct region becomes distinguishable. Consequently, its performance improves significantly at higher SNRs and eventually surpasses that of DE.

\textbf{Discussion:} The results in Figs.~\ref{fig:DEvsPSO1} and \ref{fig:DEvsPSO2} show that the relative performance of PSO and DE depends strongly on the localization scenario. In configurations where the sources are more separated, PSO can be advantageous as the problem dimension increases, since its swarm-wide information sharing helps drive the solutions toward a consistent estimate. However, this trend does not hold universally: when the sources are tightly spaced, especially at low SNR, DE becomes more robust because its mutation-selection mechanism preserves exploration more effectively and reduces the risk of premature convergence. As the SNR increases and the objective function becomes more clear, PSO benefits from its stronger exploitation capability and can again outperform DE.

\section{Comparison with Other Near-Field Localization Methods}

Grid-based subspace methods \cite{Alva2023parametric} and deep learning-based techniques \cite{Chen2025robust} are two prominent classes of approaches for near-field localization, both of which have been investigated extensively in recent years. Although these methods have shown strong performance in many prior works, they also exhibit limitations that may hinder their use in practical scenarios. We next discuss these limitations by comparing SEC with these two widely used classes of methods.

\subsection{Grid-Based Subspace Methods versus SEC}

Classical subspace methods, such as MUSIC, are well known for their high resolution and their ability to localize multiple sources. However, in near-field localization, they are typically implemented through a search over candidate angles and ranges. As a result, their accuracy depends strongly on the grid resolution. A coarse grid reduces computational complexity but introduces grid mismatch and quantization error, whereas a finer grid improves accuracy at the cost of substantial increased complexity.

SEC methods offer an attractive alternative by operating directly in the continuous parameter space and avoiding discretized grids. Rather than exhaustively evaluating a predefined grid, these methods iteratively refine a population of candidate source locations according to the chosen objective function. This makes SEC methods more flexible in high-resolution settings when exhaustive grid search becomes computationally demanding.

\subsection{Deep Learning Methods versus SEC}

Deep learning-based methods have recently attracted significant interest as they can reduce online complexity through fast inference after training, while also learning propagation or system characteristics that are difficult to model. However, these methods typically require representative labeled datasets, and their performance is often tied to the training configuration. If system parameters change, the learned model may generalize poorly, and additional retraining may be required.

By contrast, SEC methods are training-free and model-driven. They operate directly on the received measurements and the underlying signal model, without relying on labeled data or offline learning. As a result, SEC methods can be adapted more easily to new operating conditions and are particularly appealing in dynamic environments. 

\subsection{Performance Evaluation}
\begin{figure}
    \centering
    \includegraphics[width=0.9\linewidth]{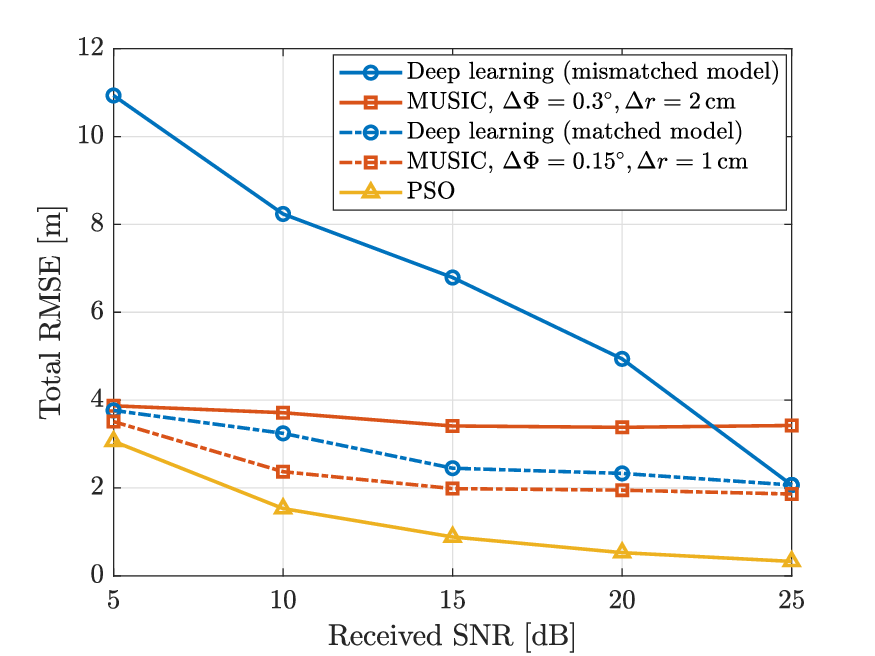}
    \caption{Performance comparison between PSO, MUSIC, and deep learning-based localization methods.  }
\label{fig:PSO_vs_benchmarks}
\end{figure}
\begin{figure}
    \centering
    \includegraphics[width=0.9\linewidth]{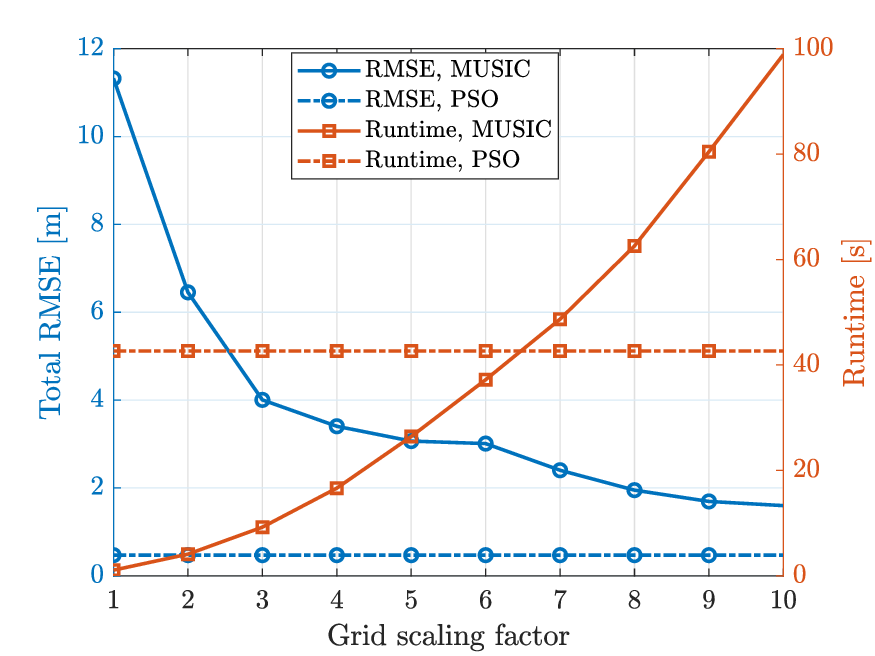}
    \caption{Performance and runtime comparison between MUSIC and PSO. }
\label{fig:MUSIC_PSO_Runtime}
\vspace{-5mm}
\end{figure}

Fig.~\ref{fig:PSO_vs_benchmarks} compares the performance of PSO with two-dimensional MUSIC and a deep learning-based method for near-field localization. The simulation parameters are the same as those used in Fig.~\ref{fig:DEvsPSO1} and three sources are considered for localization. The deep learning model consists of three convolutional layers for feature extraction, followed by several fully connected layers for regression. Specifically, convolutional kernels of sizes $5×2$, $5×1$, and $3×1$ are used, and the final output layer has $2K$ neurons ($K$ is the number of sources) to predict the location parameters.
For MUSIC, two grid resolutions are considered. In the first case, the grid sizes are $400$ points for angle and $2000$ points for range, resulting in angular and range resolution of $\Delta \phi = 0.3^\circ$ and $\Delta r = 2\,$cm, respectively. In the second case, the grid sizes are doubled, which improves the resolutions to $\Delta \phi = 0.15^\circ$ and $\Delta r = 1\,$cm, respectively. 
For deep learning, two cases are considered. In the matched case, the model is trained separately for each SNR value. In the mismatched case, a model trained at $25\,$dB is used to predict source locations at all SNR values, while the location distribution, channel model, and other simulation settings are kept unchanged. Thus, the mismatch refers only to the SNR.
The RMSE of MUSIC decreases as the grid resolution increases; however, even with a relatively fine grid, its error remains noticeably higher than that of PSO. The grid can be further refined to improve the performance of MUSIC; however, the computational cost increases rapidly with the grid resolution. This trade-off is examined in more detail in Fig.~\ref{fig:MUSIC_PSO_Runtime}.
The deep learning-based method achieves reasonable accuracy when training and testing are conducted at the same SNR. However, this matched case represents an optimistic setting, since it assumes that the operating SNR is known in advance and that a separate model can be trained for each operating condition. In practice, the SNR, channel statistics, and source configuration may vary over time, making such training impractical. The mismatched case clearly shows this limitation, as the model trained at $25\,$dB suffers a significant performance degradation at lower SNRs. This highlights the sensitivity of deep learning-based localization to the diversity of the training data.

To examine the accuracy--runtime trade-off of MUSIC, Fig.~\ref{fig:MUSIC_PSO_Runtime} depicts the RMSE and runtime at an SNR of $20\,$dB while varying the MUSIC grid resolution through a grid scaling factor $x$. Specifically, the number of angle and range grid points are set to $100x$ and $500x$, respectively. Hence, increasing $x$ from $1$ to $10$ enlarges the two-dimensional MUSIC search grid from $100\times 500$ to $1000\times 5000$ points. The RMSE and runtime of PSO are also included as reference values.
We observe that the RMSE of MUSIC decreases as the grid resolution increases, but this improvement comes at the cost of a rapidly increasing runtime. This accuracy--runtime trade-off creates scalability issues and can hinder the practical use of grid-based subspace methods such as MUSIC, which require fine grids to achieve high accuracy. By contrast, PSO achieves decent localization accuracy with a moderate runtime.

\section{Future Research Directions}
SEC-based near-field location estimation is still in its early stages, leaving considerable room for further research. In the following, we outline several directions that we believe are worth exploring.

\subsection{Finding the Number of Sources}
In many localization frameworks, the number of sources is assumed to be known in advance. In practice, however, it often must be inferred from the received data. A promising direction is to combine classical source-number estimation methods, such as akaike information criterion (AIC) or minimum description length (MDL), with SEC-based localization by running the optimizer for different assumed source numbers. Another possibility is to estimate the number of sources directly within the evolutionary process, for example, through variable-length candidate solutions or adaptive mechanisms. In both cases, new objective functions are needed whose values remain comparable across different assumed source numbers and are not biased toward any particular choice. In tracking scenarios with moving sources, temporal consistency across successive estimates can also help distinguish true sources from false detections.

\subsection{Exploiting Prior Location Information}
In some scenarios, partial prior knowledge about the source locations may be available. This can arise, for example, from previous estimates, temporal correlation in tracking scenarios, channel knowledge maps, and application-specific side information. An interesting research direction is to investigate how such prior information can be incorporated into SEC-based near-field localization, for example, through a maximum a posteriori (MAP) formulation. In this case, the objective function should combine the baseline localization objective with a prior term that reflects the available knowledge about the source locations, and the resulting optimization problem can then be solved using SEC-based methods.

\subsection{Hybrid SEC and Deep Learning Frameworks}
We have discussed the generalization issues of deep learning methods when applied to near-field localization. However, integration of these methods with SEC can be an appealing research direction. Specifically, hybrid approaches can be developed to take advantage of both the adaptability of SEC and efficiency of deep learning. For instance, deep learning can be employed to provide coarse location estimates or to narrow down the search region, while using SEC for the final optimization.  
Another possibility is to design learning models that take the SEC-based location estimates together with known system parameters, such as array configuration or propagation conditions as inputs, so that the learned model can adapt to different scenarios without requiring full retraining.

\subsection{Localization under Coherent Sources and Strong Multipath}

In this article, the sources are assumed to be uncorrelated and the channel is dominated by a line-of-sight (LoS) component with only weak non-LoS (NLoS) paths. In practice, sources may be coherent and the channel may contain strong multipath, both of which can significantly reduce localization accuracy. In such cases, objective functions based on the eigendecomposition of the sample covariance matrix are less reliable, since source correlation reduces the effective rank and makes subspace separation challenging. This motivates the development of SEC-based localization frameworks with objective functions that are more robust to source coherence and strong multipath. 

\subsection{Extension to Radar Sensing}
This article focused on localization scenarios where sources actively transmit signals to an anchor node. A natural extension is radar sensing, where targets reflect the transmitted signals instead. In this case, the signal model should account for two-way propagation, reflection coefficients, clutter, and possibly multi-static sensing. Since the objective function is central to SEC frameworks, an important direction is to design sensing-oriented objectives that capture metrics such as detection reliability, false-alarm probability, and clutter suppression.

\section{Conclusion}
This article has highlighted SEC as a promising approach for near-field localization, with the potential to overcome several limitations of existing techniques. SEC methods can operate directly in the continuous parameter space and handle complex objective landscapes without exhaustive grid search. Our comparison of PSO, as a representative swarm-based method, and DE, as a representative evolutionary method, shows that their relative advantages are scenario-dependent.
 PSO is more effective when the objective landscape is sufficiently informative, whereas DE is often more robust in ambiguous or noise-sensitive settings. We also compared PSO with baseline deep learning and subspace-based methods, and discussed the limitations of these baselines that PSO, and more generally SEC methods, can overcome.
Overall, SEC-based near-field localization remains a largely open area with substantial room for further research.
\bibliographystyle{IEEEtran}
\bibliography{refs} 
\vspace{-9mm}
\begin{IEEEbiographynophoto}{Parisa Ramezani}
received the
B.Sc. degree in electrical engineering from the
Sharif University of Technology, Iran,
in 2014, and the M.Phil. and Ph.D. degrees in
electrical and information engineering from the
University of Sydney, Australia, in 2017
and 2022, respectively. From 2022 to 2026, she was a Postdoctoral Researcher with the KTH Royal
Institute of Technology, Sweden. She is currently a chancellor's research fellow at the University of Technology Sydney, Australia. Her
research focuses on wireless communications and
signal processing.
\end{IEEEbiographynophoto}
\vspace{-11mm}
\begin{IEEEbiographynophoto}{Seyed Jalaleddin Mousavirad} received the Ph.D. degree in computer engineering (artificial intelligence) in 2016 from University of Kashan, Iran. He has held two postdoctoral research positions: first at the University of Beira Interior, Portugal, and then at Mid Sweden University, Sweden, where he worked on industrial AI projects. He is currently a Senior Lecturer at Mid Sweden University. His research interests include machine learning, pattern recognition, image processing, deep learning, and evolutionary computation. He has also been a visiting researcher at Xi’an Jiaotong-Liverpool University, China.
\end{IEEEbiographynophoto}
\vspace{-11mm}
\begin{IEEEbiographynophoto}{Mattias O'Nils} received the B.S. degree in electrical engineering from Mid Sweden University, Sundsvall, Sweden, in 1993, and the Licentiate and Ph.D. degrees in electronic systems design from the Royal Institute of Technology, Stockholm, Sweden, in 1996 and 1999,
respectively. He is currently a Professor with the Department of Computer and Electrical Engineering and leads the AI based IoT and Measurement
System Research Group, Mid Sweden University. His current research interests include implementation of AI based systems and design of AI based measurement systems.
\end{IEEEbiographynophoto}
\vspace{-11mm}
\begin{IEEEbiographynophoto}{Emil Bj\"ornson} 
 received the M.S. degree in engineering mathematics from
Lund University, Sweden, in 2007, and the Ph.D.
degree in telecommunications from the KTH
Royal Institute of Technology, Sweden, in 2011.
He is a Professor of Wireless Communication
with the KTH Royal Institute of Technology,
Stockholm, Sweden. He has a podcast and
YouTube channel called Wireless Future, has
authored four textbooks, and published much
simulation code. His research focuses on multiantenna communications and radio resource management, using methods
from communication theory, signal processing, and machine learning. He
is a Digital Futures Fellow, a Wallenberg Academy Fellow, and a Clarivate
Highly Cited Researcher.
\end{IEEEbiographynophoto}
\end{document}